\documentclass[twocolumn]{aastex63}

\usepackage{amsmath}

\received{January 26, 2020}
\revised{April 10, 2020}
\accepted{April 13, 2020}
\published{May 1, 2020}

\submitjournal{ApJL}

\shorttitle{TDE Peak-Luminosity/Decline-Rate Relationship}
\shortauthors{Hinkle et al.}

\graphicspath{{./}{figures/}}

\begin{document}

\title{Examining a Peak-Luminosity/Decline-Rate Relationship for Tidal Disruption Events}

\correspondingauthor{Jason T. Hinkle}
\email{jhinkle6@hawaii.edu}

\author[0000-0001-9668-2920]{Jason T. Hinkle}
\affiliation{Institute for Astronomy, University of Hawai`i, 2680 Woodlawn Drive, Honolulu, HI 96822, USA}

\author[0000-0001-9206-3460]{Thomas W.-S. Holoien}
\affiliation{The Observatories of the Carnegie Institution for Science, 813 Santa Barbara Street., Pasadena, CA 91101, USA}

\author[0000-0003-4631-1149]{Benjamin. J. Shappee}
\affiliation{Institute for Astronomy, University of Hawai`i, 2680 Woodlawn Drive, Honolulu, HI 96822, USA}

\author[0000-0002-4449-9152]{Katie Auchettl}
\affiliation{DARK, Niels Bohr Institute, University of Copenhagen, Lyngbyvej 2, DK-2100 Copenhagen, Denmark}
\affiliation{School of Physics, The University of Melbourne, Parkville, VIC 3010, Australia}
\affiliation{ARC Centre of Excellence for All Sky Astrophysics in 3 Dimensions (ASTRO 3D), Australia}
\affiliation{Department of Astronomy and Astrophysics, University of California, Santa Cruz, CA 95064, USA}

\author[0000-0001-6017-2961]{Christopher S. Kochanek}
\affiliation{Department of Astronomy, The Ohio State University, 140 West 18th Avenue, Columbus, OH 43210, USA}
\affiliation{Center for Cosmology and Astroparticle Physics, The Ohio State University, 191 W.~Woodruff Avenue, Columbus, OH 43210, USA}

\author{K. Z. Stanek}
\affiliation{Department of Astronomy, The Ohio State University, 140 West 18th Avenue, Columbus, OH 43210, USA}
\affiliation{Center for Cosmology and Astroparticle Physics, The Ohio State University, 191 W.~Woodruff Avenue, Columbus, OH 43210, USA}

\author[0000-0003-3490-3243]{Anna V. Payne}
\altaffiliation{NASA Fellowship Activity Fellow}
\affiliation{Institute for Astronomy, University of Hawai`i, 2680 Woodlawn Drive, Honolulu, HI 96822, USA}

\author{Todd A. Thompson}
\affiliation{Department of Astronomy, The Ohio State University, 140 West 18th Avenue, Columbus, OH 43210, USA}
\affiliation{Center for Cosmology and Astroparticle Physics, The Ohio State University, 191 W.~Woodruff Avenue, Columbus, OH 43210, USA}

\begin{abstract}
\noindent We compare the luminosity, radius, and temperature evolution of the UV/optical blackbodies for 21 well-observed tidal disruption events (TDEs), 8 of which were discovered by the All-Sky Automated Survey for Supernovae. We find that the blackbody radii generally increase prior to peak and slowly decline at late times. The blackbody temperature evolution is generally flat, with a few objects showing small-scale variations. The bolometric UV/optical luminosities generally evolve smoothly and flatten out at late times. Finally, we find an apparent correlation between the peak luminosity and the decline rate of TDEs. This relationship is strongest when comparing the peak luminosity to its decline over 40 days. A linear fit yields $\log_{10}( L_{peak}) = (44.1^{+0.1}_{-0.1}) + (1.6^{+0.4}_{-0.2})(\Delta L_{40} + 0.5)$ in cgs, where $\Delta L_{40} = \log_{10}(L_{40} /  L_{peak})$.
\end{abstract}


\section{Introduction} \label{sec:intro}
A tidal disruption event (TDE) occurs when a star passes inside the tidal radius of a supermassive black hole. The self-gravity of the star is overwhelmed by tidal forces and the star is ripped apart. This results in a luminous accretion flare \citep[e.g.,][]{rees88, evans89, phinney89, ulmer99}, with a blackbody temperature on the order of $\sim 10^5$ K \citep[e.g.,][]{lacy82, rees88, evans89, phinney89}. 

The characteristics of the observed emission from TDEs may depend on a large number of physical parameters. These include the star's impact parameter \citep[e.g.,][]{guillochon13, guillochon15}, stellar properties such as mass \citep[e.g.,][]{gallegos-garcia18, mockler19}, age \citep[e.g.,][]{gallegos-garcia18, law-smith19}, spin \citep[e.g.,][]{golightly19}, composition \citep[e.g.,][]{kochanek16a, law-smith19}, evolutionary stage \citep[e.g.,][]{macleod12}, stellar demographics \citep[e.g.,][]{kochanek16b}, the fraction of accreted stellar material \citep[e.g.,][]{metzger16, coughlin19}; the geometry of accretion \citep[e.g.,][]{kochanek94, guillochon15, dai18}; and the black hole mass and spin \citep[e.g.,][]{ulmer99, graham01, mockler19}.

Despite the large number of possibly relevant physical parameters, the UV/optical spectral energy distributions (SEDs) of TDEs are relatively well fit as blackbodies \citep[e.g.,][]{holoien14b, holoien16a, holoien16b, brown16a, hung17, holoien18a, holoien19c, leloudas19, vanvelzen20}. Generally, the effective radii increase until peak before declining monotonically. The temperature evolution is occasionally more variable, but temperatures generally remain relatively constant, or slowly increasing at late times \citep[e.g.][]{holoien19b, vanvelzen20}. The luminosity evolution is generally smooth, with only a handful of sources showing spikes or rebrightening episodes. TDE models may help explain the wide range in observed properties and variations in blackbody evolution \citep[e.g.,][]{dai18, lu20, law-smith19, ryu20}.

Pre-peak detections of TDEs are important to understanding the evolution of the UV/optical blackbody component before and after peak emission. However, TDEs are rare, with an expected frequency between $10^{-4}$ and $10^{-5} \text{ yr}^{-1}$ per galaxy \citep[e.g.,][]{vanvelzen14, holoien16a} and the discovery of TDEs before maximum brightness is challenging. Fortunately, current transient surveys like the All-Sky Automated Survey for Supernovae \citep[ASAS-SN;][]{shappee14, kochanek17}, the Asteroid Terrestrial Impact Last Alert System \citep[ATLAS;][]{tonry18}, the Panoramic Survey Telescope and Rapid Response System \citep[][]{chambers16}, and the Zwicky Transient Facility \citep[ZTF;][]{bellm19} are discovering many more TDEs, with an increasing number being discovered prior to their peak (e.g., \citealp{holoien19b, holoien19c, leloudas19, wevers19, vanvelzen20, Holoien20}, J. T. Hinkle et al. in prep.; A. V. Payne et al. in prep.).

In this Letter, we note a correlation between the peak bolometric luminosity of a TDE and its decline rate. In Section \ref{sec:sample}, we define the sample and our blackbody models of the UV/optical emission. In Section \ref{sec:L40}, we discuss the peak-luminosity/decline-rate relationship. In Section \ref{sec:discussion}, we provide discussion on this relationship. Finally, in Section \ref{sec:summary} we summarize the main results of our analysis. Throughout our analysis we assume a cosmology of $H_0$ = 69.6 km s$^{-1}$ Mpc$^{-1}$, $\Omega_{M} = 0.29$, and $\Omega_{\Lambda} = 0.71$.
 
\section{Sample and Models}
\label{sec:sample}
Because UV-optical TDEs have observed temperatures of $20,000-50,000$~K, we require UV observations to accurately determine a blackbody fit. To be included in our sample, a TDE must have been observed in the UV and optical with at least three epochs of reliable UV photometry concurrent with the optical emission in order to establish a trend in temperature. The 21 TDEs satisfying these criteria are listed in Table \ref{tab:sample}.

We divided the sources into four classes depending on how well the peak luminosity could be characterized. Class ``A'' sources have multi-band UV phototometry spanning a well-defined peak bolometric luminosity. Class ``B'' sources have ground-based data spanning a well-defined peak and multi-band UV observations within 10 days of the peak to estimate a bolometric correction for the ground-based data at peak. Class ``C'' sources have optical data spanning peak and UV more than 10 days after peak. We chose 10 days after peak to separate between Class ``B'' and ``C'' by calculating the error in peak bolometric luminosity for the Class ``A'' sources with the Swift Ultraviolet/Optical Telescope \citep[UVOT; ][]{gehrels04} photometry removed in five day bins. Ten days after peak corresponds to the point where the median of the estimated peak luminosities for each of the Class ``A'' sources differs by more than 10\% from the value found when including all available Swift photometry. Class ``D'' sources were not observed until after peak. TDEs not included in Table \ref{tab:sample} are either lacking sufficient UV and optical photometry or the UV photometry was unreliable, and thus do not meet any of the above class criteria. We include the class of each source in Table \ref{tab:sample}, along with the spectroscopic classifications introduced by \citet{vanvelzen20}. Of the 21 TDEs, only two show strong He (relative to H) lines (TDE-He), with the rest showing evidence for Bowen fluorescence emission (TDE-Bowen) or strong H lines (TDE-H).

\begin{deluxetable*}{lcccccl}[htbp!]
\tablecaption{Sample of TDEs}
\tablehead{
\colhead{Object} &
\colhead{TNS Name} &
\colhead{Class} & 
\colhead{Spectral Type} &  
\colhead{$\log_{10}( L_{peak})$} &
\colhead{$\Delta L_{40}$} &
\colhead{References} }
\startdata
ASASSN-19dj & AT2019azh & A & TDE-Bowen & $44.77 \pm 0.07$ & $-0.16 \pm 0.02$ & \citet{liu19}, \citet{vanvelzen20},  \\ & & & & & & J. T. Hinkle et al.\ (2020, in preparation) \\
ASASSN-18pg & AT2018dyb & A & TDE-Bowen & $44.35 \pm0.09$ & $-0.28 \pm 0.02$ & \citet[][]{leloudas19}, \citet[][]{Holoien20} \\
ASASSN-19bt & AT2019ahk & A & TDE-H & $44.10 \pm 0.06$ & $-0.34 \pm 0.01$ & \citet{holoien19c} \\
ZTF19abzrhgq & AT2019qiz & A & TDE-Bowen & $43.67 \pm 0.10$ & $-0.94 \pm 0.03$ & \citet{vanvelzen20} \\
Gaia-19bpt & AT2019ehz & B & TDE-H & $44.03 \pm 0.05$ & $-0.37 \pm 0.04$ & \citet{vanvelzen20} \\
ASASSN-18ul & AT2018fyk & B & TDE-Bowen & $44.78 \pm 0.07$ & $-0.31 \pm 0.06$ & \citet[][]{wevers19}, \\ & & & & & & A. V. Payne et al.\ (2020, in preparation) \\
PS18kh & AT2018zr & B & TDE-H & $43.96 \pm 0.06$ & $-0.51 \pm 0.05$ & \citet{holoien18a} \\
iPTF16fnl & AT2016fnl & B & TDE-Bowen & $43.58 \pm 0.10$ & $-1.09 \pm 0.05$ & \citet{brown18} \\
ASASSN-18zj & AT2018hyz & B & TDE-H & $44.18 \pm 0.06$ & $-0.49 \pm 0.05$ & \citet{vanvelzen20} \\
iPTF15af & \dots & C & TDE-Bowen & $43.98 \pm 0.07$ & $-0.36 \pm 0.07$ & \citet{blagorodnova19} \\
PS1-10jh & \dots & C & TDE-He & $44.45 \pm 0.08$ & $-0.40 \pm 0.08$ & \citet{gezari12b} \\
PS1-11af & \dots & C & Featureless & $43.97 \pm 0.07$ & $-0.39 \pm 0.07$ & \citet{chornock14} \\
ZTF19aapreis & AT2019dsg & C & TDE-Bowen & $44.64 \pm 0.08$ & $-0.51 \pm 0.07$ & \citet{vanvelzen20} \\
ATLAS18way & AT2018hco & C & TDE-H & $44.23 \pm 0.10$ & $-0.22 \pm 0.10$ &  \citet{vanvelzen20} \\
ZTF19abhhjcc & AT2019meg & C & TDE-H & $44.56 \pm 0.07$ & $-0.33 \pm 0.07$ & \citet{vanvelzen20} \\
PS17dhz & AT2017eqx & C & TDE-Bowen & $44.55 \pm 0.08$ & $-0.22 \pm 0.08$ & \citet{nicholl19}\\
ZTF19aabbnzo & AT2018lna & C & TDE-Bowen & $44.66 \pm 0.07$ & $-0.52 \pm 0.07$ & \citet{vanvelzen20} \\
ASASSN-15oi & \dots & D & TDE-He & $44.34_{-0.10}^{+0.33}$ & $-0.20_{-0.33}^{+0.10}$ & \citet{holoien16a} \\
ASASSN-14ae & \dots & D & TDE-H & $43.92 \pm 0.05$ & $-0.77 \pm 0.03$ & \citet{holoien14b} \\
ASASSN-14li & \dots & D & TDE-Bowen & $43.82_{-0.11}^{+0.24}$ & $-0.53_{-0.23}^{+0.09}$ & \citet{holoien16b} \\
iPTF16axa & \dots & D & TDE-Bowen & $44.04 \pm 0.05$ & $-0.42 \pm 0.05$ & \citet{hung17} \\
\enddata 
\tablecomments{The 21 TDEs studied in this Letter. Class ``A'' means that there are Swift data prior to the peak bolometric luminosity, class ``B'' means that the TDE was observed in the UV within 10 days of peak combined with ground-based photometry of the peak, class ``C'' means the TDE lacked UV observations within 10 days of peak so that bolometric corrections at peak are less reliable, and class ``D'' means that the TDE was not observed until after peak. The spectral types are taken from \citet{vanvelzen20}.}
\label{tab:sample}
\end{deluxetable*}

\begin{figure*}
\centering
 \includegraphics[width=1.0\textwidth]{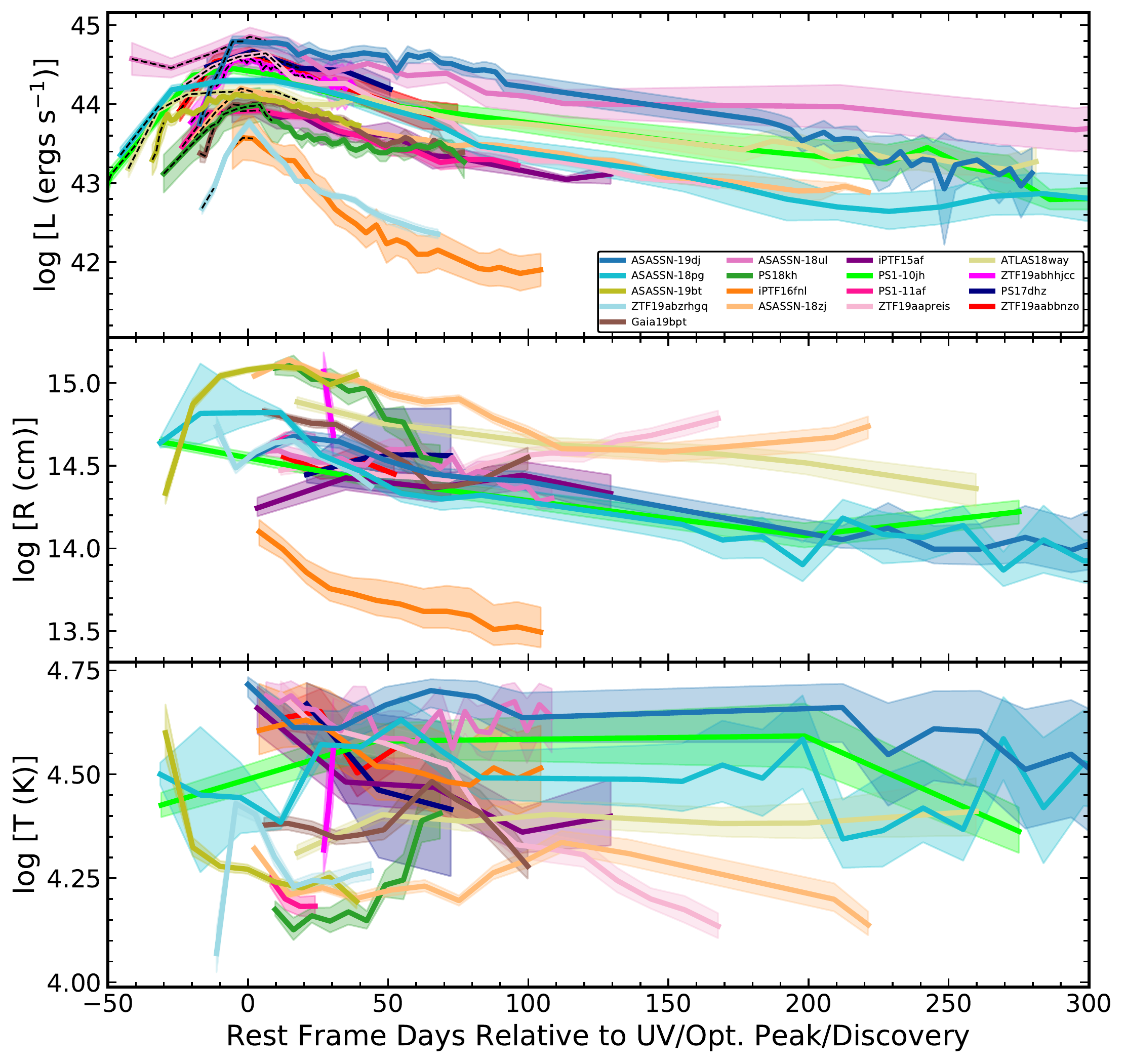}\hfill
 \caption{Evolution of the UV/optical blackbody luminosity (top panel), effective radius (middle panel), and temperature (bottom panel) for the TDEs ASASSN-19dj (blue line), ASASSN-18pg (cyan line), ASASSN-19bt (olive line), ZTF19abzrhgq (light-blue line), Gaia19bpt (brown line), ASASSN-18ul (pink line), PS18kh (green line), iPTF16fnl (orange line),  ASASSN-18zj (light-orange line), iPTF15af (purple line), PS1-10jh (lime line), PS1-11af (magenta line), ZTF19aapreis (light-pink line), ATLAS18way (light-olive line), ZTF19abhhjcc (hot-pink line),  PS17dhz (navy line), and ZTF19aabbnzo (red line). The lines are smoothed over the individual epochs by linearly interpolating to a time series with the same length as the original coverage, but with half the number of points. Time is in rest-frame days relative to the peak luminosity. Dashed lines indicate where data have been bolometrically corrected assuming the temperature from the first UV epoch.}
 \label{fig:compare}
\end{figure*}

We used Markov Chain Monte Carlo (MCMC) methods to fit a blackbody model to each epoch of UV observations for the TDEs in our sample. To keep our fits relatively unconstrained, we ran each of our blackbody fits with flat temperature priors of 10000 K $\leq$ T $\leq$ 55000 K. In Figure \ref{fig:compare}, we compare the evolution of the blackbody parameters for all but the Class ``D'' TDEs in our sample. For this figure, we smooth the lines for luminosity, radius, and temperature evolution for each TDE, by linearly interpolating to a time series with the same length as the original coverage, but with half the number of points. This allows us to compare general trends, without being overly sensitive to short-timescale variations or individual epochs of poor data quality. The time is in rest-frame days relative to the peak luminosity.

In order to accurately fit the UV photometry as a blackbody, it is necessary to subtract the emission of the host galaxy to isolate the flux from the TDE. For sources that did not have published host-subtracted photometry, we perfomed aperture photometry with an aperture that encompassed the entire host galaxy. We then fit archival photometry of the host galaxy using the Fitting and Assessment of Synthetic Templates code \citep[FAST;][]{kriek09} to obtain an SED of the host galaxy. Because in many cases the host galaxies did not have archival imaging data in the Swift UVOT filters, but rather from Galaxy Evolution Explorer \citep{martin05}, we estimated the host flux in the UVOT filters by convolving the host SED from FAST with the filter response curve for each filter to obtain fluxes matching the aperture we used for the Swift photometry. We then subtracted these synthetic host fluxes from the Swift photometry.

Our measured peak bolometric UV/optical luminosities are generally consistent given the uncertainties with values reported in the papers in Table \ref{tab:sample}. For the few objects here that are discrepant from previous reported results, there are several likely explanations. These include inconsistent host subtraction procedures, varying assumptions made when bolometrically correcting a ground-based light curve using blackbody fits, and restrictive priors when fitting for blackbody parameters. These complications make it nontrivial to directly compare our uniformly estimated peak luminosities with those in the literature.

In general, the blackbody radii increase before peak light, reaching a maximum radius near the peak or soon thereafter. After the blackbody radius peaks, there is a generally monotonic decline in size. For the TDEs with well-sampled late-time evolution (e.g., ASASSN-19dj, ASASSN-18pg, ATLAS18way), the blackbody radii continue to slowly decrease. Like \citet{vanvelzen20}, we find that the TDE-Bowen objects have smaller effective blackbody radii than the TDE-H objects.

For most of the TDEs in our sample, the blackbody temperatures stay roughly constant as they evolve. In some cases, such as ASASSN-19dj and ASASSN-18pg, there is some evolution in temperature over the first $\sim$50 days, but the general trend is flat. There are some exceptions, with ASASSN-19bt decreasing and PS18kh increasing in temperature over time. The TDE-Bowen objects are hotter than the TDE-H objects, in agreement with \citet{vanvelzen20}.

The bolometric UV/optical luminosities of the TDEs in our sample rise to peak at varying rates, and decrease roughly monotonically thereafter. The luminosity evolution of most of the objects is smooth, but some TDEs like  ASASSN-18ul, PS18kh, and ASASSN-19bt have features such as luminosity spikes or rebrightening episodes. Multiple peaks in the UV/optical luminosity could either be due to shocks caused by collisions in the debris stream \citep[e.g.,][]{gezari17} or caused by reprocessing of X-ray emission from an accretion disk \citep[e.g.,][]{wevers19, leloudas19}. The TDE-Bowen objects, with the exceptions of iPTF16fnl and ZTF19abzrhgq, are generally more luminous than the TDE-H objects, although the differences in luminosity are smaller than for the radii and temperatures.

\section{Peak-luminosity/Decline-rate Relationship} \label{sec:L40}
In Figure \ref{fig:compare}, the most luminous TDEs appear to have flatter slopes near peak, and thus decay more slowly than the less luminous TDEs. This is reminiscent of the Phillips relation for SNe Ia \citep[][]{phillips93}, except that for TDEs it is the bolometric evolution rather than the evolution in individual photometric passbands. This suggested trying to define a similar relationship for TDEs.

To determine the peak luminosity ($L_{\text{peak}}$) for the class ``A'',``B'', and ``C'' TDEs with ground-based observations prior to peak, we first bolometrically corrected the ground-based data using a linear interpolation between the UV blackbody fits before and after each ground-based observation to calculate a bolometric correction. We used the first UV epoch to define the bolometric correction for all prior epochs of ground-based observation. We then fitted a quadratic to the combined, bolometrically corrected and UV data to find the peak date and luminosity as well as the associated uncertainties. For the Class ``D'' sources, we simply took the maximum bolometric luminosity and uncertainty as the peak luminosity and corresponding uncertainty. 

We incorporated uncertainties in distance into the uncertainties on peak luminosity. These come from both uncertainties in the value of the Hubble constant and the spread in observed peculiar velocities. We summed the statistical and systematic errors from \citet{freedman20} in quadrature to yield an error of $1.9 \text{ km s}^{-1}\text{ Mpc}^{-1}$ on $H_{0}$. From the observed distribution of peculiar velocities in the nearby universe, we assume a representative spread of $500 \text{ km s}^{-1}$ \citep[e.g.,][]{tully16}.

\begin{figure}
\centering
 \includegraphics[width=0.48\textwidth]{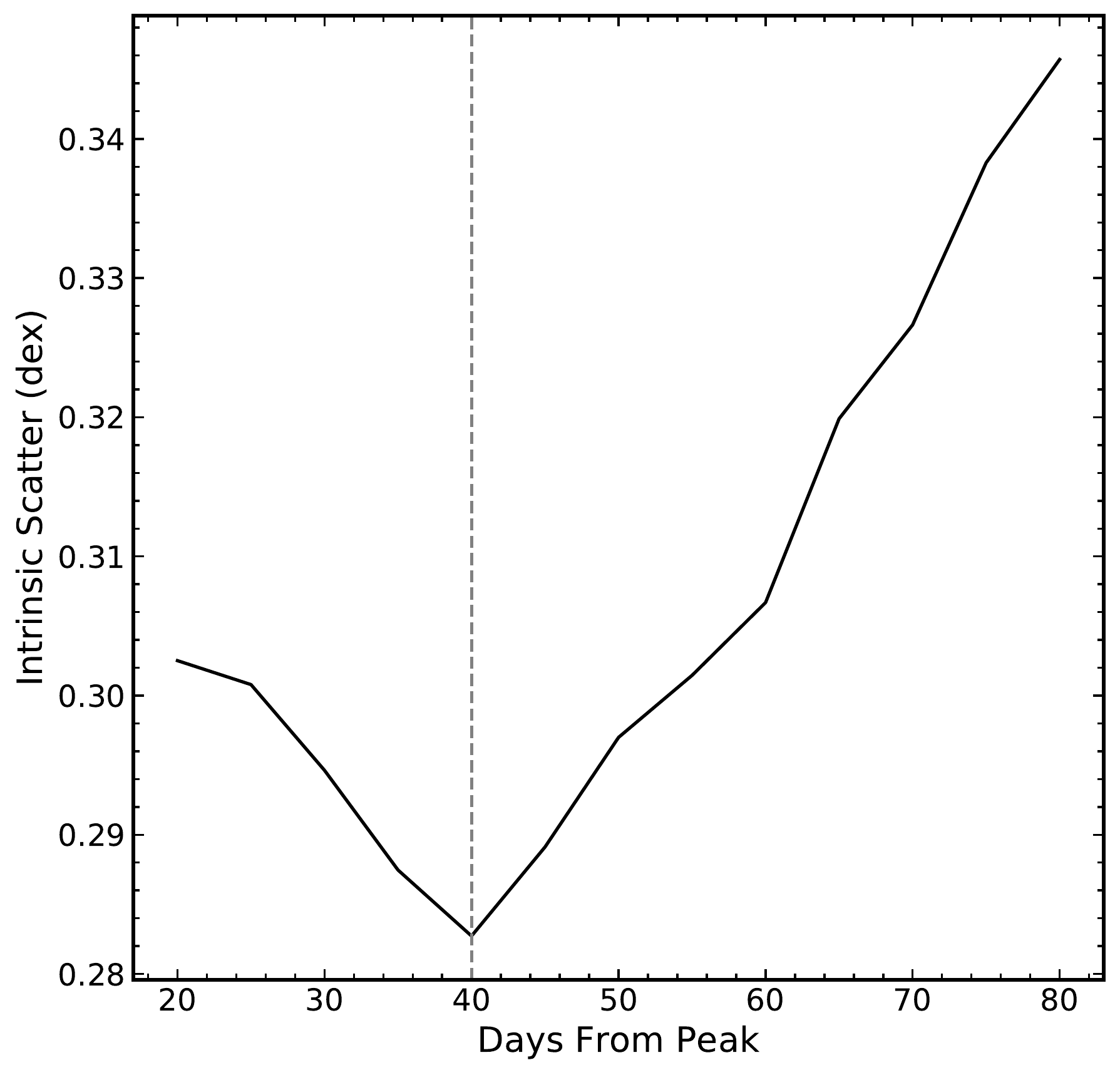}\hfill
 \caption{Fitted intrinsic scatter in the relationship between $\log_{10}( L_{peak})$ and $\Delta L_{N}$ where we sampled $N$ in steps of 5 days from 20 days after peak until 80 days after peak. The dashed gray line indicates the chosen value of 40 days at the minimum intrinsic scatter of 0.284 dex.}
 \label{fig:int_scatter}
\end{figure}

We quantified the decline rate of TDEs using a similar parameterization to $\Delta\text{M}_{15}$ in the Phillips relation with

\begin{figure*}
\centering
 \includegraphics[width=1.0\textwidth]{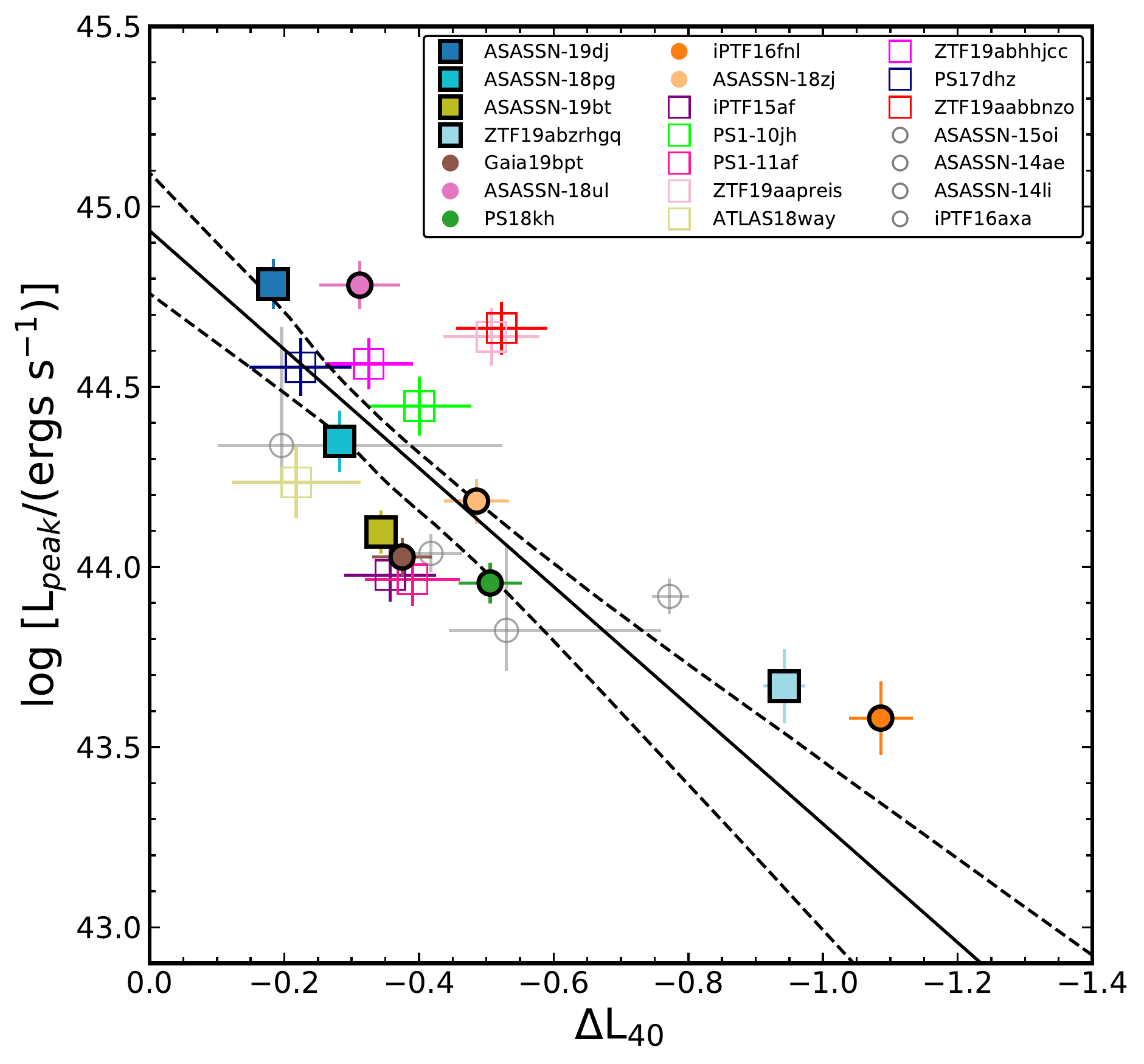}\hfill
 \caption{Peak bolometric UV/optical luminosity as compared to the decline rate $\Delta L_{40} = $ log$_{10}(L_{40}/ L_{peak})$, where $L_{40}$ is the luminosity of the TDE at 40 days after peak. The objects and colors are the same as in Figure \ref{fig:compare}. Filled squares with a black border are our ``A'' sample, filled circles with a black border are the ``B'' TDEs, open squares are the ``C'' TDEs, and the gray open circles are the ``D'' TDEs. The solid black line is the line of best fit and the dashed black lines are plus/minus one sigma from the best-fit line. We have excluded the Class ``D'' objects in our MCMC fit.}
 \label{fig:L40}
\end{figure*}

\begin{equation}
\Delta L_{N} = \text{log}_{10}\left(\frac{L_{N}}{L_{peak}}\right) = \text{log}_{10}(L_{N}) - \text{log}_{10}(L_{peak})
\end{equation}

\noindent where $L_{N}$ is the luminosity of the TDE at $N$ days after peak. We calculated $\Delta L_{N}$ by fitting a line to the data within 10 days of $N$, centered on $N$, and taking the intercept minus the peak luminosity. Based on our analysis of the change in peak luminosity for the Class ``A'' sources, we added 10\% and 15\% errors in quadrature to the peak luminosity of the Class ``B'' and Class ``C'' objects respectively. We fitted the relationship between $ L_{peak}$ and $\Delta L_{N}$ using the method of \citet{kelly07}, which is a linear MCMC fit including an intrinsic scatter parameter. To determine the optimal $N$ at which to measure the decline rate, we examined the fitted intrinsic scatter of the relationship between $ L_{peak}$ and $\Delta L_{N}$ for $20 < N < 80$ in 5 day steps. As seen in Figure \ref{fig:int_scatter}, the minimum intrinsic scatter is $\sim$0.29 dex at 40 days. For the rest of the study we adopt $N=40$ days, although this may need to be modified as more TDEs are analyzed.

\begin{figure*}
 \includegraphics[width=0.48\textwidth]{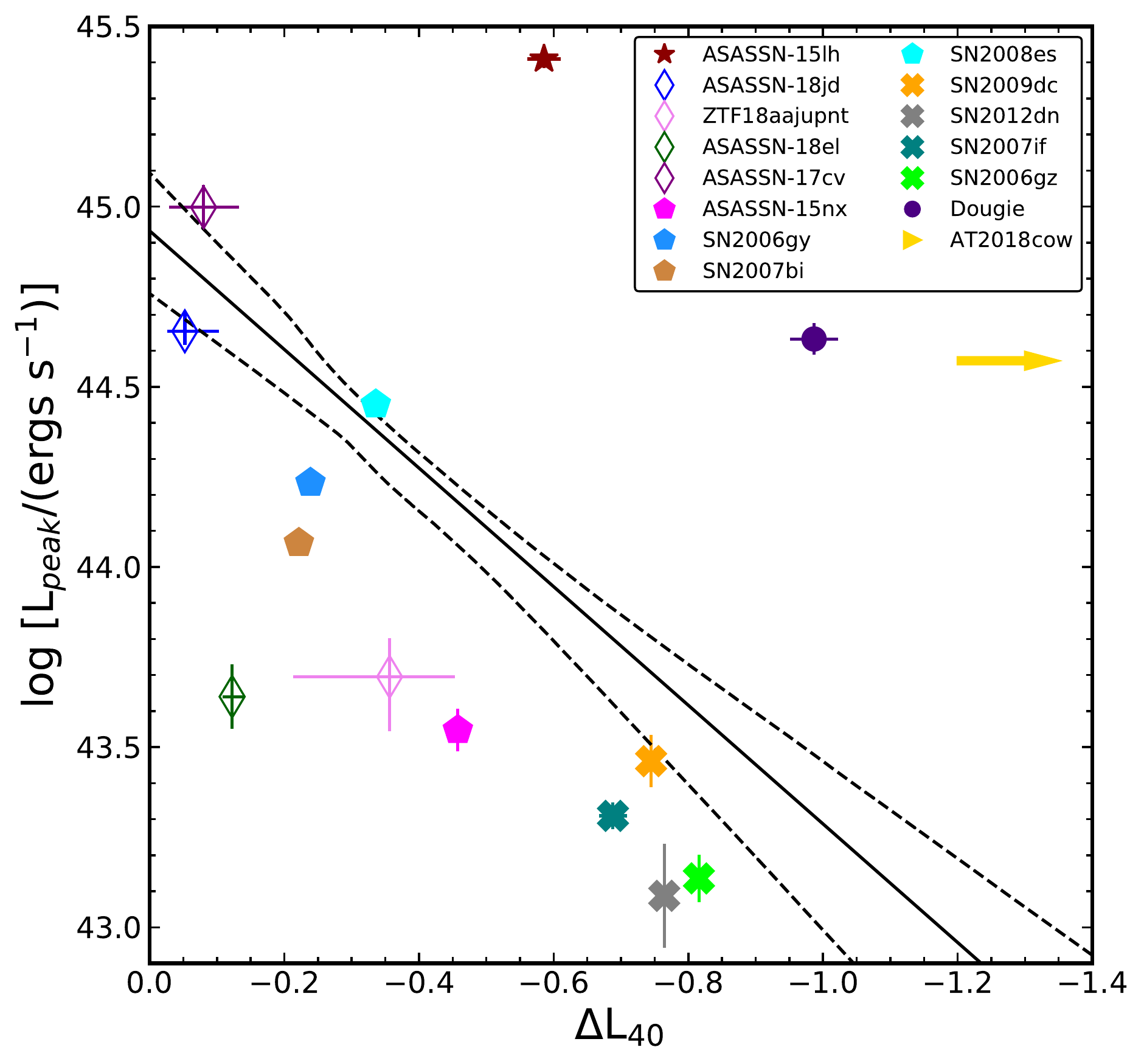}\hfill
 \includegraphics[width=0.48\textwidth]{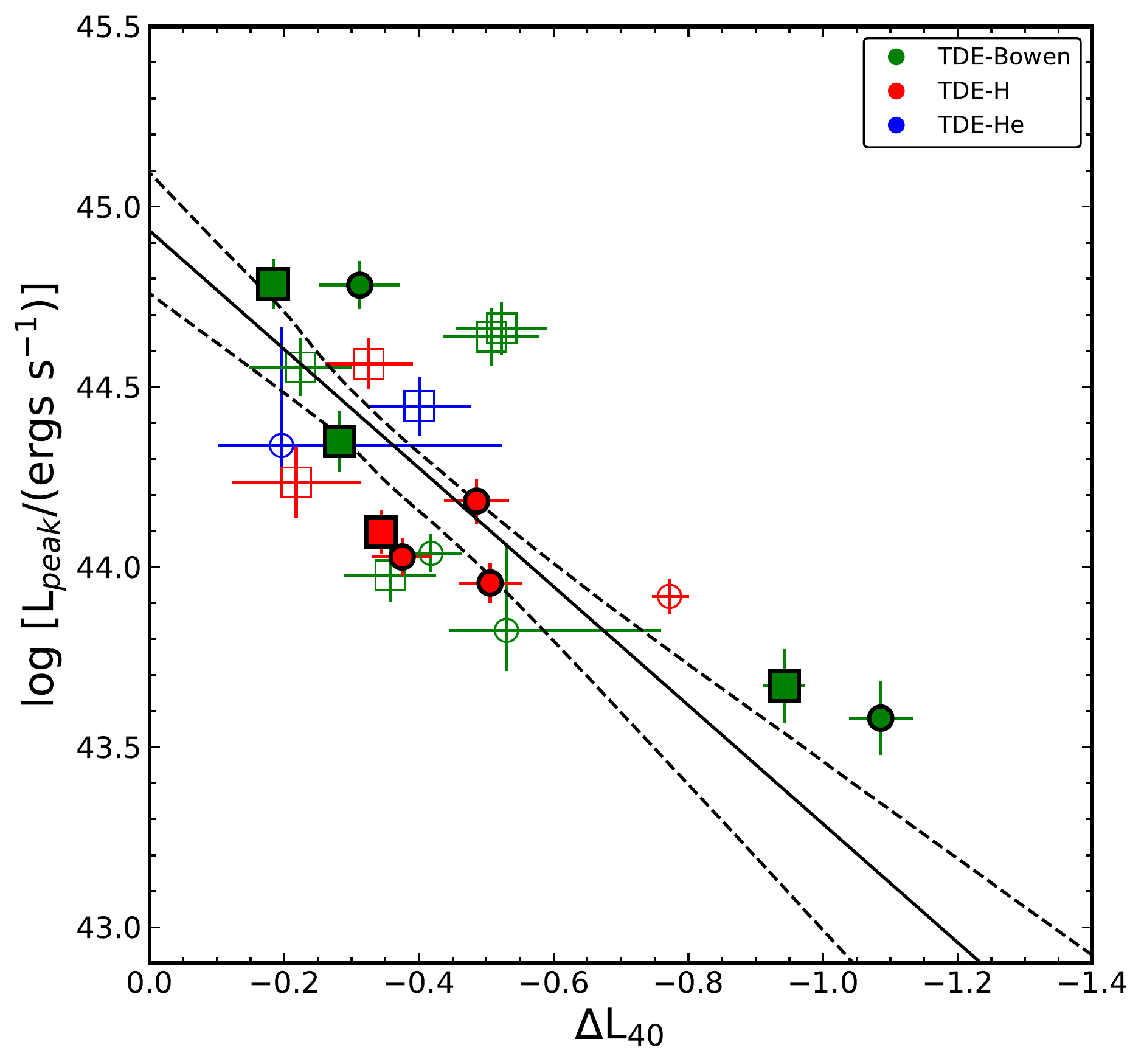}\hfill
 \caption{Left panel: other luminous transients compared to the best-fit relationship from Figure \ref{fig:L40}. The other transients are ASASSN-15lh (a SLSN or TDE; red star), ASASSN-18jd (an AGN or TDE; blue diamond), ZTF18aajupnt (a changing-look LINER; violet diamond), ASASSN-18el (a changing-look AGN; green diamond), ASASSN-17cv (an AGN flare; purple diamond), ASASSN-15nx (a luminous SN II; pink pentagon), SN2006gy (an SLSN II; light-blue pentagon), SN2007bi (an SLSN II; brown pentagon), SN2008es (an SLSN II; aqua pentagon), SN2009dc (a super-Chandrasekhar SN Ia; orange ``X''), SN2012dn (a super-Chandrasekhar SN Ia; gray ``X''), SN2007if (a super-Chandrasekhar SN Ia; teal ``X''), SN2006gz (a super-Chandrasekhar SN Ia; lime ``X''), and Dougie (a fast UV transient; indigo circle) The fast UV transient AT2018cow \citep{perley19} lies off the right of the diagram, with the peak luminosity indicated here by a gold arrow. Here the filled symbols have data spanning the peak luminosity, and the unfilled symbols take the maximum value as the peak luminosity. Right panel: TDEs from Figure \ref{fig:L40}, color-coded with respect to their spectral types \citep{vanvelzen20}, with the exception of PS1-11af whose spectra are featureless. Green objects are TDE-Bowen, red are TDE-H, and blue are TDE-He.}
 \label{fig:L40_other}
\end{figure*}

The resulting correlation between $L_{peak}$ and $\Delta L_{40}$ is shown in Figure \ref{fig:L40}. Using the procedure of \citet{kelly07}, we fit the ``A'', ``B'', and ``C'' objects with a linear function, and the best fit is

\begin{equation}
\begin{split}
\log_{10}(L_{peak} / (\text{erg s}^{-1})) = (44.1^{+0.1}_{-0.1}) + \\ (1.6^{+0.4}_{-0.2})(\Delta L_{40} + 0.5)
\end{split}
\end{equation}

\noindent where $-0.5$ is the approximate mean of the values of $\Delta L_{40}$. We include this offset in the linear fits because it makes the uncertainties in the two parameters essentially uncorrelated. If we fit only the Class ``A'' and ``B'' sources we find

\begin{equation}
\begin{split}
\log_{10}( L_{peak} / (\text{erg s}^{-1})) = (44.1^{+0.1}_{-0.4}) + \\ (1.6^{+1.9}_{-0.4})(\Delta L_{40} + 0.5)
\end{split}
\end{equation}

\noindent which is consistent with the fit to the Class ``A'', ``B'', and ``C'' sources. The uncertainties we quote on our best-fit parameters are estimated using bootstrap resampling with 10,000 iterations. The reduced $\chi^2$ values are 4.2 (14 dof) and 5.2 (5 dof) respectively, if we do not allow for any intrinsic scatter. When included, the estimated intrinsic scatter in peak luminosities is $0.29^{+0.23}_{-0.17}$ dex, or roughly a factor of 2 in peak luminosity. While this relationship is analogous to that seen for SNe Ia \citep{phillips93}, the scatter is larger than that of the Type Ia supernovae (e.g., \citealp{folatelli10}). Given their rarity, similar peak optical magnitudes, and larger scatter, TDEs are unlikely to be competitive distance indicators. 

\section{Discussion} \label{sec:discussion}
We can gain some insight into the physical meaning of this relation by examining the scaling under the assumption that the mass of the black hole is the dominant driver of luminosities and timescales. The change in luminosity over time $\Delta t$ is of order $\delta L = \Delta t (L_{peak} / t_d)$, where $t_d$ is some decay timescale. This means that $\Delta L_N = \log_{10}\left({\Delta t_N / t_d} \right)$, so the scaling relation in Equation 2 is that $L_{peak} \propto t_d^{-a}$ where $a$ is the slope of the relation. If $t_d$ is related to the standard fall back time, $t_d \propto t_{fb} \propto M_{BH}^{1/2}$, then the peak luminosity scales with mass as $L_{peak} \propto M_{BH}^{-a/2} \propto M_{BH}^{-0.8\pm 0.2}$ given our parameters. This is close to estimates that the peak accretion rate relative to Eddington is $\propto M_{BH}^{-3/2}$ or $\dot{M}_{peak} \propto M_{BH}^{-1/2}$ with $L_{peak} \propto \dot{M}_{peak}.$ \citep[e.g.,][]{metzger16, kochanek16b, mockler19, ryu20}. 

The right panel of Figure \ref{fig:L40_other} shows our sample and fit, with the points color-coded by their spectroscopic classification \citep{vanvelzen20}. There seem to be few trends between the spectral type of a TDE and its position on the peak-luminosity/decline-rate diagram. Both the most and least luminous sources are TDE-Bowen objects, with the TDE-H objects falling in between. Accordingly, neither spectral type of TDE appears to decay faster than the other. The TDE-Bowen objects appear to have slightly larger scatter about the correlation.

The correlation between peak luminosity and decline rate may also provide a diagnostic for whether sources are actually TDEs. The left panel of Figure \ref{fig:L40_other} shows the correlation found for TDEs along with several other nuclear transients. ASASSN-18el \citep{trakhenbrot19b} is a changing-look AGN with no present arguments in favor of it being a TDE, and we see that it lies far off the relation. ASASSN-18jd shows similarities to both TDEs and nuclear flares that look different from normal AGN variability \citep{neustadt20}, but here we see that it is consistent with the relation, albeit with the caveat that it is a class ``D'' source. \citet{dong16} and \citet{bersten16} classify ASASSN-15lh as a Type I superluminous supernova (SLSN), while \citet{leloudas16} classify it as a TDE, even though no well-studied TDEs show similar spectroscopic properties or evolution. Here we see that it lies far off the correlation, supporting the SLSN classification. The AGN flare ASASSN-17cv \citep{trakhenbrot19a} lies roughly one sigma from the relationship while the changing-look LINER ZTF18aajupnt \citep{frederick19} lies far below the trend. We also include four super-Chandrasekhar Type Ia SNe from \citet{taubenberger19} and find that only one of the four is consistent with the relationship, with most falling below. The luminous SN II ASASSN-15nx \citep{bose18a} also falls below the relationship. We additionally include three superluminous SNe II and a fast UV transient from \citet{vinko15}. SN2008es lies on the relationship, SN2006gy is consistent with the position of some of the TDEs in our sample, and SN2007bi lies below the relationship. The fast UV transient ``Dougie'' \citep{vinko15} lies far above the relationship, fading faster than TDEs with similar peak luminosities. Another fast UV transient, AT2018cow \citep{perley19}, is luminous ($\log_{10}( L_{peak}) = 44.6 \pm 0.04$), but fades much faster than any of the other transients shown here ($\Delta L_{40} = -2.7 \pm 0.04$).

\section{Summary} \label{sec:summary}
From the TDE blackbody fits shown in Figure \ref{fig:compare} we find the following trends:

\begin{itemize}
\item The blackbody radii generally are largest near peak and monotonically decline as time passes. At late times ($\gtrsim$ 200 days), the blackbody radius continues to decrease slowly. The TDE-Bowen objects generally have smaller effective blackbody radii than the TDE-H objects.

\item Most of the TDEs have roughly constant temperatures with some small-scale variations. Only a handful of objects show large-scale increases or decreases in their temperature. The TDE-Bowen objects are generally hotter than the TDE-H objects.

\item The luminosities of the TDEs generally evolve smoothly, but some exhibit multiple spikes in luminosity. At late times, the luminosities of the TDEs flatten out.

\item As can be seen in Figure \ref{fig:compare}, more luminous TDEs fade more slowly. The correlation is strongest when we use the decline over 40 days, with a slope of $\log_{10}(L_{peak}) \sim (1.6^{+0.4}_{-0.2})(\Delta L_{40} + 0.5)$.

\item As shown in the left panel of Figure \ref{fig:L40_other}, the relationship between peak luminosity and decline rate for TDEs may prove useful for selecting TDEs from other luminous, blue, nuclear transients.
\end{itemize}

The biggest shortcoming of the present sample is that roughly half of the sources were not observed within 10 days of peak in the UV (the class ``C'' and ``D'' sources). This is a further reason to emphasize the early discovery and classification of TDEs, which fortunately seems to be increasingly common due to a growing number of transient surveys covering large fractions of the sky at high cadence and with significant survey overlaps. To further test this correlation observationally, it will be necessary to discover more TDEs early in their evolution and obtain high signal-to-noise ratio Swift UVOT follow-up photometry, as accurately fitting the blackbody components of TDEs requires UV coverage. For example, \citet{vanvelzen20} do not find a relationship between the peak luminosity and monochromatic decay timescales (ZTF $g$ or $r$ band). \citet{vanvelzen20} assume a single, mean temperature for the early-time evolution of each TDE. In our fits to the available data, this can misestimate the temperature at peak by $\sim$10\%, leading to $\sim$40\% errors in $\log_{10}( L_{peak})$. This further indicates the need for UV follow-up to constrain the bolometric decline. If future TDEs confirm this correlation, theoretical simulations and models of the UV and optical emission from TDEs must be able explain this trend.

\acknowledgments
We thank the referee Suvi Gezari for helpful comments and suggestions that have improved the quality of this manuscript. We thank Michael Tucker and Aaron Do for helpful comments on the manuscript, Gagandeep Anand for useful discussions that improved our error analysis, and Jack Neustadt for sharing blackbody fits for ASASSN-18jd. We also thank Sjoert van Velzen for helpful discussion on archival photometry used in this work.

We thank the Las Cumbres Observatory and its staff for its continuing support of the ASAS-SN project. ASAS-SN is supported by the Gordon and Betty Moore Foundation through grant GBMF5490 to the Ohio State University, and NSF grants AST-1515927 and AST-1908570. Development of ASAS-SN has been supported by NSF grant AST-0908816, the Mt. Cuba Astronomical Foundation, the Center for Cosmology and AstroParticle Physics at the Ohio State University,  the Chinese Academy of Sciences South America Center for Astronomy (CAS- SACA), the Villum Foundation, and George Skestos. 

B.J.S, K.Z.S, and C.S.K are supported by NSF grants AST-1515927, AST-1814440, and AST-1908570. B.J.S is also supported by NSF grants AST-1920392 and AST-1911074. K.A.A is supported by the Danish National Research Foundation (DNRF132). T.A.T is supported in part by NASA grant 80NSSC20K0531.

Parts of this research were supported by the Australian Research Council Centre of Excellence for All Sky Astrophysics in 3 Dimensions (ASTRO 3D), through project number CE170100013.

\facilities{ASAS-SN, Swift(UVOT)}
\software{linmix \citep{kelly07}}

\bibliography{bibliography}{}
\bibliographystyle{aasjournal}

\end{document}